\documentclass[journal]{IEEEtran}
\usepackage{cite}
\usepackage{mathrsfs}
\usepackage{amsmath,amssymb,amsfonts}
\usepackage{graphicx}
\usepackage{subfigure}
\usepackage{algorithmic}
\usepackage{textcomp}
\usepackage{xcolor}
\usepackage{amsmath,bm}
\usepackage{multirow}
\usepackage{algorithm}
\usepackage{booktabs}
\usepackage{epstopdf}

\ifCLASSINFOpdf
\else
\fi
\begin{document}
	\title{CSI Feedback with Model-Driven Deep Learning of Massive MIMO Systems}
	\author{Jianhua Guo,
		Lei Wang,
		Feng Li,
		and Jiang Xue, ~\IEEEmembership{Senior Member,~IEEE}% <-this % stops a space
		\thanks{J. Guo, L. Wang, F. Li and J. Xue are with School of mathematics and statistics, Xi'an Jiaotong University, Xi'an, 710049, China (e-mail: \{jhguo0525, wl\_simple\}@stu.xjtu.edu.cn, lifeng53@huawei.com and x.jiang@xjtu.edu.cn)}
	}

	\maketitle
	
	\begin{abstract}
		In order to achieve reliable communication with a high data rate of massive multiple-input multiple-output (MIMO) systems in frequency division duplex (FDD) mode, the estimated channel state information (CSI) at the receiver needs to be fed back to the transmitter. However, the feedback overhead becomes exorbitant with the increasing number of antennas. In this paper, a two stages low rank (TSLR) CSI feedback scheme for millimeter wave (mmWave) massive MIMO systems is proposed to reduce the feedback overhead based on model-driven deep learning. Besides, we design a deep iterative neural network, named FISTA-Net, by unfolding the fast iterative shrinkage thresholding algorithm (FISTA) to achieve more efficient CSI feedback. Moreover, a shrinkage thresholding network (ST-Net) is designed in FISTA-Net based on the attention mechanism, which can choose the threshold adaptively. Simulation results show that the proposed TSLR CSI feedback scheme and FISTA-Net outperform the existing algorithms in various scenarios.
	\end{abstract}
	

	% \fi
	\IEEEpeerreviewmaketitle

	\section{Introduction}
	\IEEEPARstart{A}{s} a key technology for the 5th-Generation (5G) wireless communication systems, massive multiple-input multiple-output (MIMO) is one of the currently attractive technologies for future wireless access \cite{6736761}. There is a prerequisite for using massive MIMO, which is the base station (BS) needs to acquire channel state information (CSI) of the downlink channel. In time division duplex (TDD) mode, BS can obtain downlink CSI from uplink channel based on the channel reciprocity \cite{6654952}. In frequency division duplex (FDD) mode, the reciprocity is no longer hold and the downlink CSI needs to be estimated at the user equipment (UE) based on pilot and fed back to the BS. However, the enormous amount of feedback information make the feedback overhead become unbearable in the massive MIMO systems because of the surge in the number of antennas.
	
    The conventional method with pre-defined codebook to quantize CSI into a codeword and reconstruct CSI according to the index of the codeword in codebook \cite{8379322}, which can not meet the system requirements due to the feedback overhead linearly with the number of antennas. Based on the channel sparsity caused by direction-of-arrivals are mainly concentrated only a few of directions, compressed sensing (CS) technology is exploited widely to recover downlink CSI from a low-dimensional compressed information \cite{6816089}. Although CS-based methods can reduce the CSI feedback overhead, it usually transforms the problem of CSI reconstruction into an optimization problem solved by iterative algorithms with huge consumption and computational resources. Meanwhile, their performance relies heavily on the manual setting hyper-parameters, such as step size and threshold, which brings great uncertainty for CSI feedback. In addition, these methods only exploit the channel sparsity, but ignore the property of low rank \cite{8122055}, \cite{8812960}, such as in millimeter wave (mmWave) systems, which should also be considered to further improve the performance of CSI feedback. 
	
	In recent years, deep learning (DL), in virtue of the powerful learning ability, has been successfully applied in the physical layer of wireless communications to improve the performance, such as MIMO detection \cite{9159940}, channel estimation \cite{8640815} and CSI feedback which exploits convolutional neural networks as the encoder and decoder to compress and reconstruct CSI respectively \cite{8322184, 8972904, 9076084, 9279228, 9495802}. Although DL-based methods have a respectable performance for CSI feedback, they are data-driven modes and usually regarded as a black box. The relationship between network structure and its performance is vague, which resulting in the difficulty for designing and explaining the network. To overcome these issues, the model-driven deep learning is proposed in \cite{10.1093/nsr/nwx099}, which unfolding a traditional algorithm for a special task as a deep neural network with powerful learning ability, easy designing and interpretable network structure, such as ADMM-Net \cite{DBLP:journals/corr/YangSLX17}, ISTA-Net \cite{DBLP:journals/corr/ZhangG17b} and ISTA-Net+ \cite{8578294}.
	
	In this paper, we propose a two stages low rank (TSLR) CSI feedback scheme for low rank mmWave channel, with a model-driven neural network by unfolding the fast iterative shrinkage thresholding algorithm (FISTA), named FISTA-Net. Specifically, the low rank channel matrix is decomposed into two parts which can be expressed linearly with each other. These two parts are compressed at UE and fed back to BS respectively. After receiving the feedback information, BS exploits the FISTA-Net and the linear relationship between decomposed parts to reconstruct the two parts of the low rank channel matrix separately, and stitch them into CSI matrix based on the order of decomposition. 
	
	The main contributions of this paper are listed below.
	\begin{itemize}
		\item TSLR CSI feedback scheme is designed based on the feedback of low rank mmWave channel. With the help of TSLR, we can transform the problem of CSI reconstruction into two lightweight reconstruction problems and provide an efficient initial method to obtain excellent performance. 
		
		\item We unfold the FISTA for solving the sparse reconstruction problem as a model-driven deep iterative neural network, namely, FISTA-Net, which retains the powerful learning ability and has an easy designing and interpretable structure.
		
		\item In FISTA-Net, we developed a shrinkage thresholding network (ST-Net) based on the attention mechanism, which can obtain a set of thresholds based on the input features of network adaptively.
	\end{itemize}

	\section{System Model}
	In this work, we consider a mmWave massive MIMO system with $N_{t}$ and $N_{r}$ antennas at BS and UE respectively. The mmWave channel using a geometric channel model can be expressed as \cite{8812960}
	%********************************Equation(1)********************************
	\setcounter{equation}{0}
	\begin{equation}
	\bm{H}=\sum_{l=1}^{L}\alpha_{l}\bm{a}_r\left(\theta_{l}\right)\bm{a}^H_t\left(\phi_{l}\right),
	\end{equation}
	where $\bm{H}\in \mathbb{C}^{N_{r}\times N_{t}}$, $L$ is the number of propagation, $\alpha_{l}\in \mathcal{CN}(0, 1/2)$ denotes the complex gain of the $l$th path, $\bm{a}_r\in \mathbb{C}^{N_{r}}$ and $\bm{a}_t\in \mathbb{C}^{N_{t}}$ represent array response vector at UE and BS respectively, $\theta_{l}$ and $\phi_{l}$ denote the physical angle of arrival (AoA) and angle of departure (AoD), which generated from the laplace distribution respectively, and $(\cdot)^H$ represents conjugate transpose.
	
	We assume that UE has known the perfect CSI, and the total number of elements for CSI feedback without compressing is $N_{r}N_{t}$, which is huge and unbearable for massive MIMO systems. To reduce the feedback overhead by CS algorithms based on the channel sparsity, we transform the mmWave channel into beam space domain as \cite{8122055}
	%********************************Equation(2)********************************
	\setcounter{equation}{1}
	\begin{equation}
	\bm{H}=\bm{D}_r\bm{H}_{sl}\bm{D}^{H}_t,
	\end{equation}
	where $\bm{D}_r \in \mathbb{C}^{N_r \times N_r}$ and $\bm{D}_t \in \mathbb{C}^{N_t \times N_t}$ are unitary matrices based on the Discrete Fourier Transform (DFT), and $\bm{H}_{sl}\in \mathbb{C}^{N_{r}\times N_{t}}$ is a sparse matrix contains only a few of virtual channel gains with high amplitude. In addition, \cite{8122055}, \cite{8812960} indicate the mmWave channel has a low rank structure in the case of angular spreads result from scattering clusters and the rank of channel $\bm{H}$ is
	%********************************Equation(3)********************************
	\setcounter{equation}{2}
	\begin{equation}
	rank(\bm{H}) \leqslant \sum_{l=1}^{L}rank\left(\bm{a}_r\left(\theta_{l}\right)\bm{a}^H_t\left(\phi_{l}\right)\right)=L < N_r,
	\end{equation}
	which reveals that the matrix $\bm{H}_{sl}$ has both sparsity and low rank structure \cite{8812960}.
	
	To overcome the difficulty of the complex operation in neural network, we rewrite the complex matrix $\bm{H}_{sl}$ as
	%********************************Equation(4)********************************
	\setcounter{equation}{3}
	\begin{equation}
	\bm{\widetilde{H}} = \left[\mathfrak{Re}(\bm{H}_{sl}^T), \mathfrak{Im}(\bm{H}_{sl}^T)\right]^T,
	\end{equation}
	where $\bm{\widetilde{H}} \in \mathbb{R}^{2N_{r} \times N_{t}}$ is still a low rank sparse matrix and the rank of $\bm{\widetilde{H}}$ is denoted as $rank(\bm{\widetilde{H}}) = R$, $(\cdot)^T$ represents transpose, $\mathfrak{Re}(\cdot)$ and $\mathfrak{Im}(\cdot)$ denote the real and imaginary part of a complex number, respectively.
	
	\section{TSLR CSI Feedback}
	In this section, we will present our proposed TSLR CSI feedback scheme for mmWave massive MIMO systems. 
	
	Based on the low rank structure of the mmWave channel, we regard all column vectors of low rank channel matrix $\bm{\widetilde{H}}$ as a column space and decompose $\bm{\widetilde{H}}$ as
	%********************************Equation(5)********************************
	\setcounter{equation}{4}
	\begin{equation}
	\bm{\widetilde{H}}=\left[\bm{\widetilde{H}}_1, \bm{\widetilde{H}}_2\right],
	\end{equation}
	where $\bm{\widetilde{H}}_1\in \mathbb{R}^{2N_{r}\times R}$ is a basis of the column space matrix $\bm{\widetilde{H}}$, $\bm{\widetilde{H}}_2\in \mathbb{R}^{2N_{r}\times (N_{t}-R)}$ is the residual part of the column space and can be expressed linearly by $\bm{\widetilde{H}}_1$ as $\bm{\widetilde{H}}_2=\bm{\widetilde{H}}_1\bm{B}$, where $\bm{B}\in \mathbb{R}^{R\times \left({N_{t}-R}\right)}$ is the coefficient matrix. We denote $\bm{h}_1=vec(\bm{\widetilde{H}}_1)$, $\bm{h}_2=vec(\bm{\widetilde{H}}_2)=(\bm{I} \otimes \widetilde{\bm{H}}_1)vec(\bm{B}) \triangleq \bm{M}\bm{b}$, where $\bm{M}=\bm{I} \otimes \widetilde{\bm{H}}_1$, $\bm{b} = vec(\bm{B})$, $vec(\cdot)$ and $\otimes$ represent the vectorized function and kronecker product respectively.

	\begin{figure}[t]
		\centerline{\includegraphics[width=3.2in,height=1.5in]{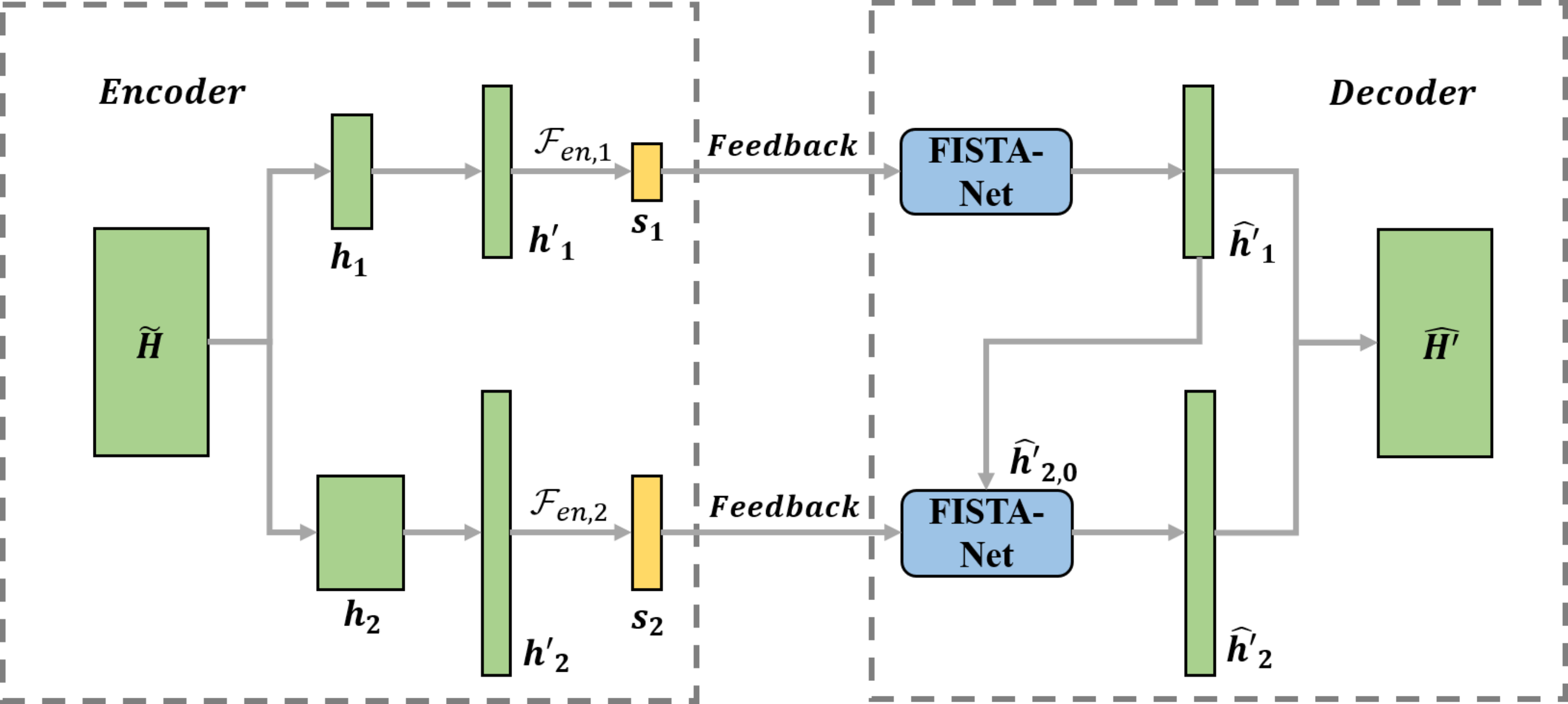}}
		\caption{Design of TSLR CSI feedback scheme. The left module is an encoder at the UE to compress $\bm{h}'_1$ and $\bm{h}'_2$ of $\bm{H}$. Correspondingly, the right module is a decoder at the BS to reconstruct the CSI matrix $\bm{H}$ by FISTA-Nets from the received compressed informations.}
		\label{Fig1}
	\end{figure}
	
	In this paper, we design a TSLR mechanism for low rank CSI feedback. As shown in Fig. \ref{Fig1}, we adopt two fully connected layers without bias $\mathcal{F}_{en, 1}$ and $\mathcal{F}_{en, 2}$ as the encoder to compress $\bm{h}_1$ and $\bm{h}_2$ to low-dimensional codewords $\bm{s}_1$ and $\bm{s}_2$, which will be fed back to BS respectively, i.e.,
	%********************************Equation(6)********************************
	\begin{subequations}
		\begin{equation}
		\bm{s}_1=\bm{W}_{en,1}\bm{h}_1 + \bm{n}_1, 
		\end{equation}
		\begin{equation}
		\bm{s}_2=\bm{W}_{en,2}\bm{h}_2 + \bm{n}_2=\bm{W}_{en,2}\bm{M}\bm{b} + \bm{n}_2, 
		\end{equation}
	\end{subequations}
	where $\bm{W}_{en, 1}$ and $\bm{W}_{en, 2}$ are the weight matrices of $\mathcal{F}_{en, 1}$ and $\mathcal{F}_{en, 2}$ respectively, $\bm{n}_{i},i=1,2$ are additive white Gaussian noise (AWGN). Although we assume the perfect CSI is known at UE, the consideration of noise during the feedback link is necessary \cite{9076084}. In decoder, the sparsity of $\bm{h}_{i},i=1,2$ can be exploited to recovery CSI and the reconstruction problem can be written as
	%********************************Equation(7)********************************
	\setcounter{equation}{6}
	\begin{equation}
	\min_{\bm{h}_i}\frac{1}{2}\left\|\bm{s}_i - \bm{W}_{en,i}\bm{h}_i\right\|^2_2 + \tau_i
	\|\mathcal{T}_{i}(\bm{h}_i)\|_1, \quad i = 1, 2,
	\end{equation}
	where $\mathcal{T}_{i}(\cdot), i=1,2$ represent the sparse transform and $\tau_i, i=1,2$ are penalty parameters. In this way, the problem of reconstructing $\bm{\widetilde{H}}$ is converted to two lightweight sparse reconstruction problems. To solve the above problems, we unfold the FISTA to a deep iterative neural network, named FISTA-Net, which will be introduced in section  $\rm \uppercase\expandafter{\romannumeral4}$, as the decoder for reconstructing $\bm{h}_1$ and $\bm{h}_2$ respectively. 
	
	\begin{figure*}[t]
		\centering
		\centerline{\includegraphics[width=5.9in,height=2.9in]{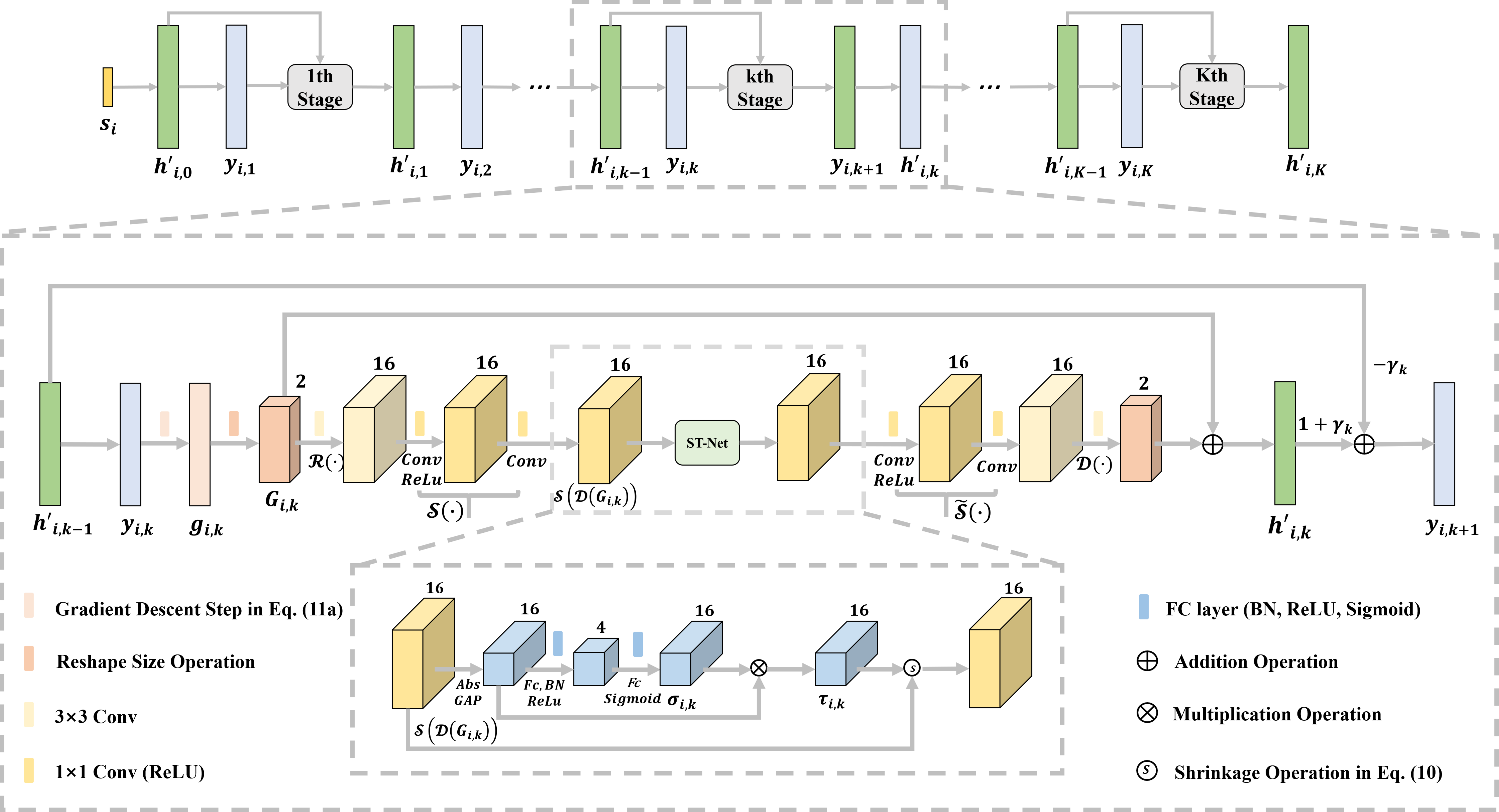}}
		\caption{The structure of the FISTA-Net by unfolding FISTA. The rectangles and cubes represent 2-dim vectors and 4-dim tensors and the numbers above the cubes represent the channels of the tensors.}
		\label{Fig2}
	\end{figure*}
	
	In general, the initial inputs of a iterative algorithm are usually set as zero, random value or from the least square (LS) estimator. In the case of low rank channel, we consider a more efficient initial value of $\bm{h}_2$ according to the linear relationship between $\bm{h}_1$ and $\bm{h}_2$ in TSLR. Specifically, when the estimated vector $\bm{\hat{h}}_1$ of $\bm{h}_1$ has been derived, we can obtain a rough solution of $\bm{b}$ by LS estimator as
	%********************************Equation(8)********************************
	\setcounter{equation}{7}
	\begin{equation}
	\bm{\hat{b}}= (\bm{W}_{en,2}\bm{\hat{M}})^{\dagger}\bm{s}_2,
	\end{equation}
     where $\bm{\hat{M}}=(I \otimes \bm{\bm{\hat{H}}})_1$, $\bm{\bm{\hat{H}}}_1$ is the reshaped original size matrix of $\bm{\hat{h}}_1$, $\dagger$ is the pseudo inverse operation. Then, the initial value of $\bm{h}_2$ can be expressed as $\bm{h}_{2,0}= \bm{\hat{M}}\bm{\hat{b}}$. In this way, all of the prior information of $\bm{\hat{h}}_1$ are considered, so that the $\bm{h}_{2, 0}$ is more close to the $\bm{h}_2$ and the FISTA-Net for $\bm{h}_2$ converges faster and has a better performance.

	\section{FISTA-Net}
	
	In this section, we review the basic iteration steps of FISTA. Later the design details of FISTA-Net will be introduced.
	%fast iterative shrinkage thresholding algorithm (FISTA). 

	\subsection{FISTA}
	As a classical algorithm for solving sparse recovery problems, FISTA is the enhanced version of iterative shrinkage thresholding algorithm (ISTA) by nesterov acceleration and widely used to solve various CS reconstruction problems with excellent performance \cite{beck2009fast}. For solving problem (7), we start with $\bm{y}_{i,1}=\bm{h}_{i,0}$, and the major update steps of FISTA in $k$th iteration are as follows
	%********************************Equation(9)********************************
	%********************************Equation(12)********************************
	\begin{subequations}
		\begin{equation}
		\bm{g}_{i,k} = \bm{y}_{i,k}-\eta_i \bm{W}^T_{en,i}\left(\bm{W}_{en,i} \bm{y}_{i,k} -   \bm{s}_i\right),
		\end{equation}
		\begin{equation}
		\bm{h}_{i, k} = \mathcal{T}^{-1}_{i}(soft\left(\mathcal{T}_{i}(\bm{g}_{i,k}),\tau_i \right)),
		\end{equation}
		\begin{equation}
		\bm{y}_{i, k+1} = \bm{h}_{i, k} + \frac{t_k-1}{t_{k+1}}\left(\bm{h}_{i, k} - \bm{h}_{i, k-1}\right),
		\end{equation}
	\end{subequations}
	where $k$ is the iteration index of FISTA, $\eta_i$ is the step size for the $i$th ($i=1,2$) reconstruction problem, $t_k$ denotes shrinkage pseudo coefficient and can be updated by $t_{k+1} = \frac{1+\sqrt{1+4t^2_k}}{2}$, $t_1=1$, and $soft\left(\bm{g}, \tau \right)$ is the shrinkage function in the following form
	%********************************Equation(11)********************************
	\setcounter{equation}{9}
	\begin{equation}
	soft\left(\bm{g}, \tau \right) = sign\left(\bm{g}\right)max\left\{|\bm{g}|-\tau, 0\right\},
	\end{equation}
	where $sign\left(\cdot \right)$ is symbolic function. Eq. (9a), Eq. (9b), and Eq. (9c) show the process of gradient descent, thresholding shrinkage, and nesterov accelerated gradient, respectively. There are some hyper-parameters in FISTA that need to be set through experiences manually, such as $\eta_i$, $\tau_i$ and $t_1$, which bring great uncertainty for CSI feedback if they are set inappropriately.
	
	\subsection{Unfolding FISTA to FISTA-Net}
	To exploit the powerful learning ability of DL and reduce the uncertainty for CSI reconstruction, we unfold all iterative steps of FISTA to deep neural network layers and design an iterative neural network, named FISTA-Net, which includes $K$ stages corresponding to $K$ iterations of FISTA. As depicted in Fig. \ref{Fig2}, we use learnable parameters and ST-Net to replace the manual setting hyper-parameters in the FISTA (step size, threshold and shrinkage cofficient), which can choose the optimal parameters during network training. In the $k$th stage, the operations of unfolding FISTA to FISTA-Net mainly include the following aspects.
	\begin{itemize}
		\item[1)] \textbf{Parameters \ Learnability:}
		To avoid the bias of manual setting, we take the step size $\eta_i$ and shrinkage coefficient $\frac{t_k-1}{t_{k+1}}$ as the learnable parameters $\eta_{i, k}$ and $\gamma_{i, k}$ of FISTA-Net respectively. To increase network capacity while retaining the structure of FISTA, we allow $\eta_{i,k}$ and $\gamma_{i,k}$ to be different at each iteration stage. Therefore, $\bm{g}_{i,k}$ and $\bm{y}_{i, k+1}$ in Eq. (9) can be expressed as
		%********************************Equation(14)********************************
		\begin{subequations}
			\begin{equation}
			\bm{g}_{i,k} = \bm{y}_{i,k}-\eta_{k} \bm{W}^T_{en,i}\left(\bm{W}_{en,i} \bm{y}_{i,k} -   \bm{s}_i\right),
			\end{equation}
			\begin{equation}
			\bm{y}_{i, k+1} = \bm{h}'_{i, k} + \gamma_{k} \left(\bm{h}_{i, k} - \bm{h}_{i, k-1}\right).
			\end{equation}
		\end{subequations}
		
		After the operation of Eq. (11a), we reshape the vector $\bm{g}_{i, k}$ to a $N_r \times W_i \times 2$ matrix, which can be regarded as a 2-channel image $\bm{G}_{i, k}$ to facilitate the operations of subsequent convolution layers.
		
		\item[2)] \textbf{Sparse\ Transform\ and\ Residual\ Learning:}
		Although the mmWave channel has the sparsity, sometimes it may be not enough. In order to get a more sparse representation of CSI, a non-linear transform $\mathcal{S}(\cdot)$ including two $1 \times 1$ convolution layers without biases separated by a rectified linear unit (ReLU)  and the inverse of the non-linear transform $\widetilde{\mathcal{S}}(\cdot)$ with the same structure proposed by \cite{DBLP:journals/corr/ZhangG17b} are applied in FISTA-Net to replace $\mathcal{T}_i(\cdot)$ and $\mathcal{T}^{-1}_i(\cdot)$, respectively. Meanwhile, to get richer feature information, similar to \cite{8578294}, two $3 \times 3$ convolution layers without biases $\mathcal{R}(\cdot)$ and $\mathcal{D}(\cdot)$, are placed before sparse transformation $\mathcal{S}(\cdot)$ and after the inverse of sparse transformation $\widetilde{\mathcal{S}}(\cdot)$ separately. In addition, to avoid the exploding gradient problem caused by stacking many hidden layers, a shortcut structure is designed in FISTA-Net, which add the $\bm{G}_{i,k}$ to the feature map after the convolution layer $\mathcal{D}(\cdot)$. Thus $\bm{h}_{i, k}$ in Eq. (9b) can be rewritten as
		%*******************************Equation(15)*******************************
		\setcounter{equation}{11}
		\begin{equation}
		\bm{h}_{i,k} = \bm{G}_{i, k} + \mathcal{D}\left(\widetilde{\mathcal{S}}\left(
		soft\left(\mathcal{S}\left(\mathcal{R}\left(\bm{G}_{i,k}\right)\right), \tau_i \right)\right)\right).
		\end{equation}

		\item[3)] \textbf{ST-Net}:
		The selection of threshold in Eq. (9b) has a great influence on the performance of FISTA. Inspired by \cite{8850096}, we design a sub-network based on the attention mechanism to search threshold $\tau_i$. The structure of ST-Net is shown at the bottom of Fig. \ref{Fig2}. The absolute value of sparse transformed feature map $|\mathcal{S}\left(\mathcal{R}\left(\bm{G}_{k,i}\right)\right)|$ is first reduced into a 16-dimensional vector by global average pooling (GAP). Then, two fully connected layers $\mathcal{F}_1$ and $\mathcal{F}_2$ with batch normaliztion (BN) have 4 and 16 neurons, respectively, are used to generate a scaling factor $\bm{\sigma}_{k, i}$, whose activation functions are ReLU and sigmoid respectively. The $k$th iteration threshold $\bm{\tau}_{k,i}$ can be obtained by multiplying the GAP reduced-dimensional vector and scaling vector $\bm{\sigma}_{k,i}$. Noted the threshold $\bm{\tau}_{k,i}$, generated by $\bm{G}_{k,i}$, is a 16-dimensional vector and varies at different stages, we can obtain a set of thresholds for each channel based on the characteristics of the input feature map adaptively. Then we proceed to the shrinkage function for $\mathcal{S}\left(\mathcal{R}\left(\bm{G}_{k,i}\right)\right)$ according to Eq. (10).
		
	\end{itemize}
	
	We denote the trainable parameter set in FISTA-Net by $\boldsymbol{\varTheta}$, includes the step size $\eta_{i, k}$, shrinkage coefficient $\gamma_{i, k}$ in Eq. (11), the parameters of convolution layers in Eq. (12), and the parameters of two fully connected layers in ST-Net. That is, the trainable parameter set of FISTA-Net in $K$ stages is $\boldsymbol{\varTheta} = \left\{ \eta_{k}, \gamma_{k}, \mathcal{S}_{k}, \mathcal{\widetilde{S}}_{k}, \mathcal{R}_{k}, \mathcal{D}_{k}, \mathcal{F}_{1,k}, \mathcal{F}_{2,k} \right\}^K_{k=1}	$.
	Besides, two fully connected layers without bias $\left\{ \mathcal{F}_{en, 1}, \mathcal{F}_{en, 2} \right\}$ to compress CSI in encoder also need to be learned.
	
	Due to FISTA-Net contains $K$ stages with the inverse transform $\widetilde{\mathcal{S}}$ of $\mathcal{S}$, the loss function designed in the training process can be written as
	%*******************************Equation(16)*******************************
	\setcounter{equation}{12}
	\begin{align}
	%		\begin{split}
	\mathcal{L}(\boldsymbol{\varTheta}) & = \mathcal{L}_{mse} + \mu \mathcal{L}_{iteration} + \zeta \mathcal{L}_{symmetry} \notag \\
	& = \left\| \bm{\hat{h}}_{i, K} - \bm{h}_i\right\|^2_2 +  
	\mu \sum_{k=1}^{K} \left\| \bm{\hat{h}}_{i, k} - \bm{h}_i\right\|^2_2 \notag \\
	& +  \zeta \sum_{k=1}^{K} \left\| 	\widetilde{\mathcal{S}}\left(\mathcal{S}\left(\bm{\hat{h}}_{i, k}\right) - \bm{h}_i\right) \right\|^2_2,
	%		\end{split}
	\end{align}
	where $\mu$ and $\zeta$ are two trade-off parameters. The $\mathcal{L}_{mse}$ is mean squared error (MSE) to measure the difference between the ground truth and the output of FISTA-Net, which is the key component of the loss function. The $\mathcal{L}_{iteration}$ is the sum of error between ground truth and iterative value in each stage of the FISTA-Net, which helps to derive estimates close to the ground truth at each stage. The $\mathcal{L}_{symmetry}$ is to make $S \circ \widetilde{S} = \bm{I}$ hold as well as possible.
	
	\section{Simulation Results}
	
	This section provides details of the experimental and comparative results in different scenarios.
	
	\subsection{Experiment Setting}
	We use the uniform linear array (ULA) with $N_r=N_t=16$ antennas at UE and BS to generate mmWave channel $\bm{H}$ at 90 GHz. The number of propagation $L$ is set to 2 and the AoA and AoD generated from the Laplace distribution with standard deviation $50^{\circ}$. After transforming to beam space domain and real field, the rank of low rank sparse matrix $\bm{\widetilde{H}}$ is $R=4$. The training set and test set contain 10,066 and 4,315 samples respectively. The experiment is implemented in Tensorflow on the NVDIA GeForce RTX 2080. We use Adam optimizer to train FISTA-Net with $K=20$ stages and the epochs, batch size, learning rate and two trade-off parameters $\mu$ and $\zeta$ in loss function are set to 300, 32, 0.001, 0.01 and 0.01 respectively. The FISTA-Nets for $\bm{h}_1$ and $\bm{h}_2$ have the same structure and hyper-parameters except the initial values which zeros are adopted for $\bm{h}_1$ and the TSLR getting initial value of $\bm{h}_2$ from $\bm{\hat{h}}_1$. 
	
	\subsection{Performance Analysis}
	To evaluate the accuracy of the CSI reconstructed by different methods, we use the normalized mean square error (NMSE) as the performance metric, which is defined as 
	%*******************************Equation(16)*******************************
	\setcounter{equation}{13}
	\begin{align}
        NMSE = \mathbb{E}\left\{
        	\frac{\|\bm{\hat{H}}-\bm{H}\|^2_2}{\|\bm{H}\|^2_2}
        	\right\},
	\end{align}
% 	$NMSE = \mathbb{E}\left\{
% 	10\lg\frac{\|\bm{\hat{H}}-\bm{H}\|^2_2}{\|\bm{H}\|^2_2}
% 	\right\}$, 
	where $\bm{\hat{H}}$ is the reconstructed channel matrix by reshaping size and splicing $\bm{\hat{h}}_1$ and $\bm{\hat{h}}_2$ according to Eq. (5), and $\bm{H}$ is the truth channel matrix.

% 	\begin{figure}[t]
% 		\begin{minipage}[t]{0.48\linewidth}
% 			\centerline{\includegraphics[width=1.85in,height=1.45in]{initial5_in_dB.pdf}}
% 			\caption{NMSE vs SNRs for $\bm{h}_2$.}
% 			\label{Fig3}
% 		\end{minipage}
% 		\begin{minipage}[t]{0.48\linewidth}
% 			\centerline{\includegraphics[width=1.85in,height=1.45in]{compare3_in_dBwith.pdf}}
% 			\caption{NMSE vs SNRs for $\bm{H}$.}
% 			\label{Fig4}
% 		\end{minipage}
% 	\end{figure}
	\begin{figure}[htpb]
		\centerline{\includegraphics[width=3.6in,height=2.65in]{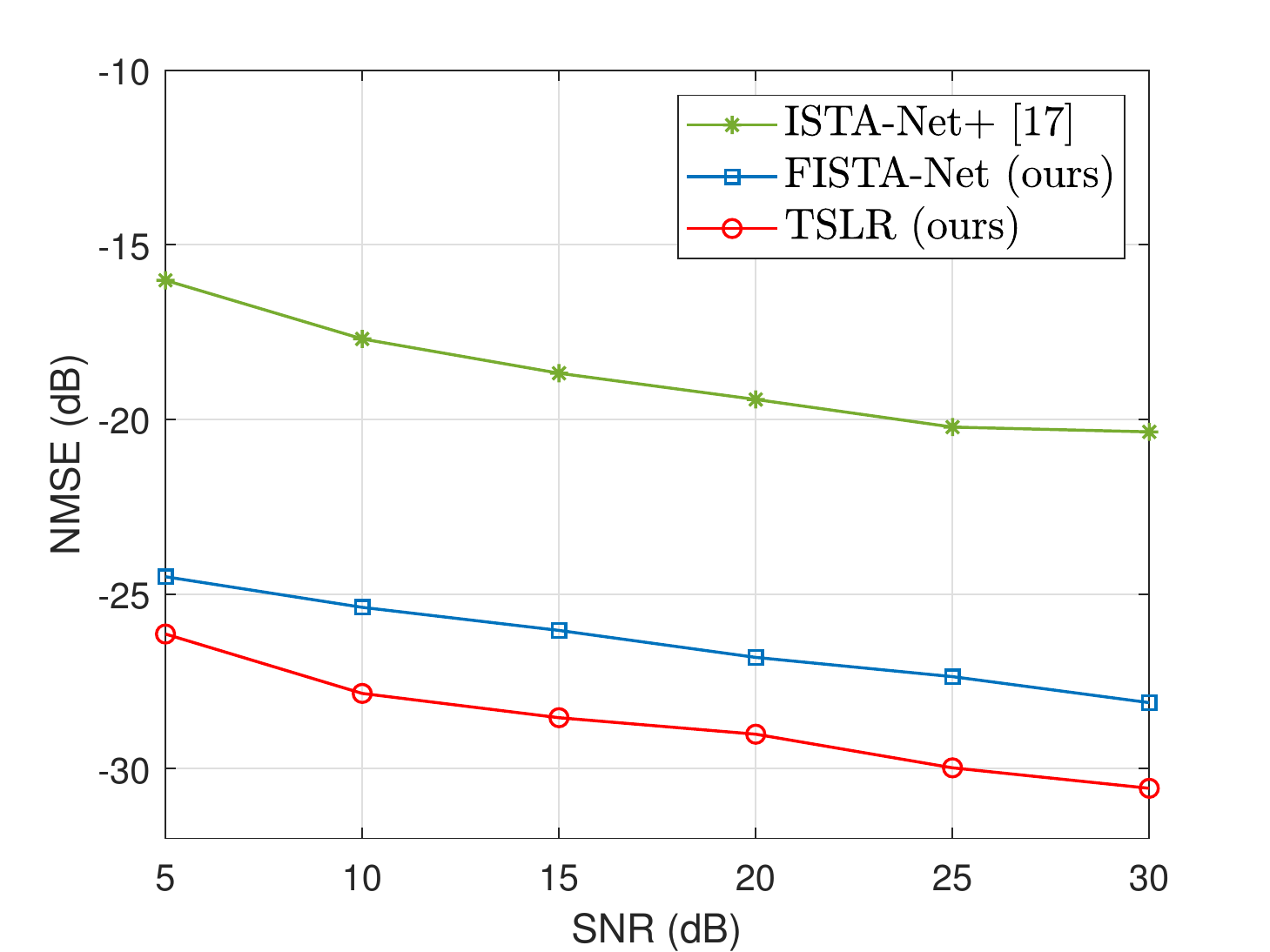}}
		\caption{NMSE (dB) vs SNR for the reconstruction of $\bm{h}_2$.}
		\label{Fig3}
	\end{figure}

	\begin{figure}[htpb]
		\centerline{\includegraphics[width=3.6in,height=2.65in]{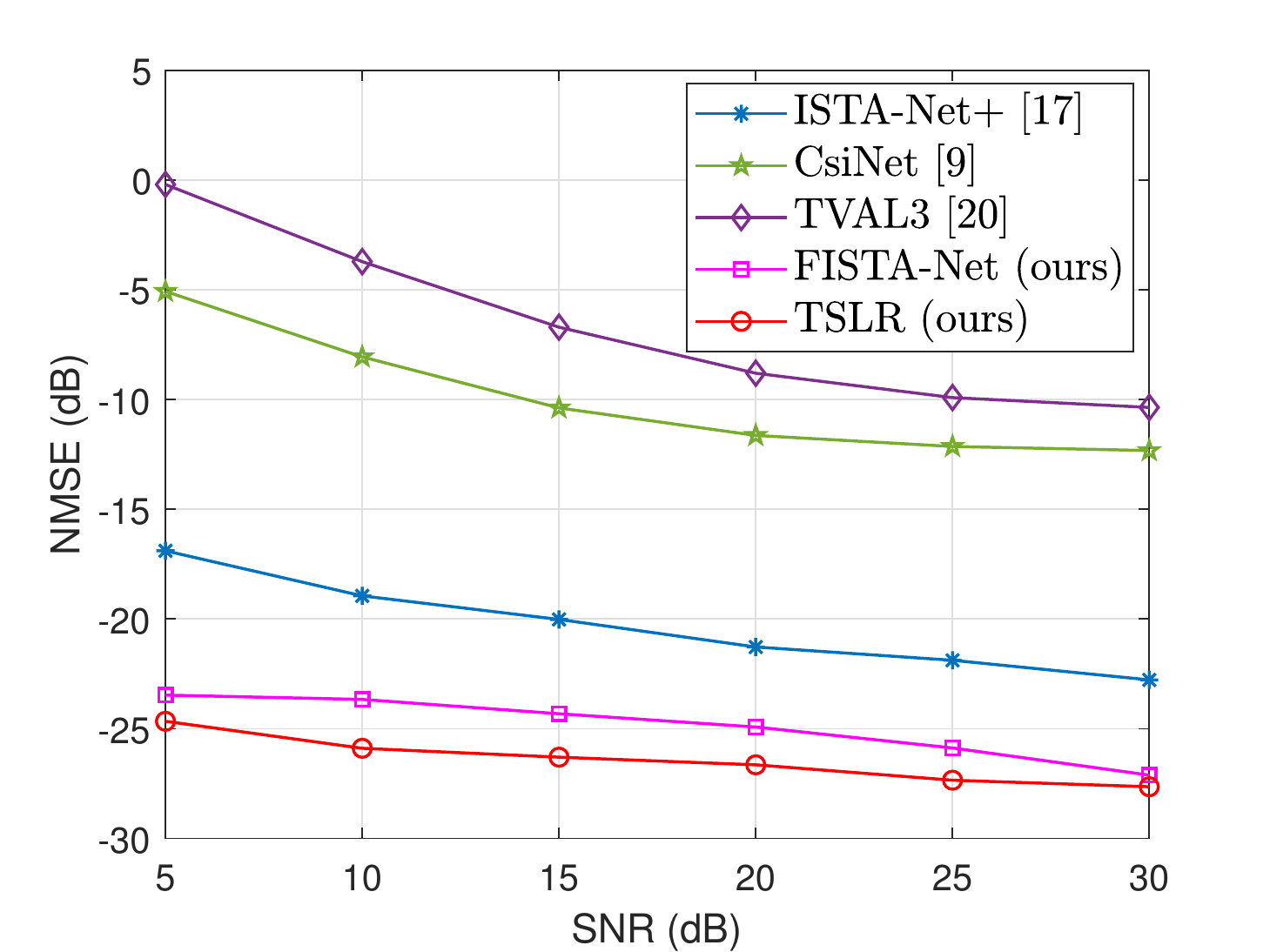}}
		\caption{NMSE (dB) vs SNR for the reconstruction of the CSI matrix $\bm{\widetilde{H}}$.} 
		\label{Fig4}
	\end{figure}
	
	When the compression ratio (CR) of $\bm{h}_1$ and $\bm{h}_2$ are both 1/4, Fig. \ref{Fig3} depicts the performance comparison of reconstruction $\bm{h}_2$ by different methods with various initial value at different SNR, where ISTA-Net+ adopts LS method to obtain initial value of $\bm{h}_{2}$ and FISTA-Net adopts zeros as the initial value. We can see from Fig. \ref{Fig3} that FISTA-Net shows better performance than ISTA-Net+, which indicates the FISTA-Net is effective. Besides, we proposed TSLR by getting the initial value of $\bm{h}_2$ by $\bm{\hat{h}}_1$ can further improve the performance of FISTA-Net for reconstructing $\bm{h}_2$.

	% 	\begin{figure}[htpb]
	% 		\centerline{\includegraphics[width=3.6in,height=2.8in]{compare3.eps}}
	% 		\caption{NMSE performance comparison for $\bm{\hat{H}}$ between differnet methods at different SNRs when CR is 1/4.}
	% 		\label{Fig4}
	% 	\end{figure}

	As shown in Fig. \ref{Fig4}, four classic CSI feedback algorithms, including TVAL3 \cite{li2013tval3}, CsiNet \cite{8322184}, ISTA-Net+ \cite{8578294} and FISTA-Net are applied to compare TSLR at different SNRs, where the CR is still 1/4. From the Fig. \ref{Fig4}, it is clear that TSLR demonstrated excellent performance at all SNR. For other CSI reconstruction methods, TVAL3 performs the worst. Although CsiNet performs better than TVAL3, is performance is worse than the model-driven methods. Compared to the ISTA-Net+ and FISTA-Net, TSLR has a distinct improvement which further verified the effectiveness of FISTA-Net and the advantage of the TSLR scheme compared with reconstructing the whole CSI matrix $\bm{\widetilde{H}}$ directly.

	\subsection{Performance in OFDM System}
	At the current CSI feedback research, most of the methods consider the single-cell downlink massive MIMO system with $N_t=32$ antennas at BS and a single antenna at UE without feedback noise \cite{8322184}. The orthogonal frequency division multiplexing (OFDM) with $\widetilde{N}_c=1024$ sub-carriers is adopted in the system. By 2-dimensional DFT, the spatial-frequency channel $\bm{H}=[\bm{h}'_1,...,\bm{h}'_{\widetilde{N}	_c}]^H \in \mathbb{C}^{\widetilde{N}_c \times N_t}$ also holds the sparstiy with the first $N_c$ rows contain large value, denoted by $\bm{H}'$, in the angular-delay domain \cite{8322184}.

	The channel is generated by COST 2100 \cite{6393523} channel model at 5.3 GHz indoor scenario and 300 MHz outdoor scenario with ULA. The training and testing channel dataset contain 100,000 and 20,000 samples, respectively. We use a fully connected layer as the encoder to compress and feedback $\bm{H}'$ and a FISTA-Net as the decoder to reconstruct $\bm{H}'$ in different CRs. The loss function is given by Eq. (13) and the Adam optimizer is used to train FISTA-Net includes $K=20$ stages with 300 epochs and 64 batch size. Moreover, the initial learning rate is 1e-3 and multiply 0.1 per 100 epochs.

	We compare FISTA-Net with FISTA \cite{beck2009fast}, CsiNet \cite{8322184} and its enhanced version CsiNet+ \cite{8972904}. The NMSE (dB) and the complexity including the number of trainable parameters and multiply-accumulate (MACC) operations in the encoder and decoder under different CRs and scenarios are given in Table \ref{tab1}, where the best results are marked as the bold font.
    \begin{table}[htpb]
    \centering
    \caption{NMSE in $\rm dB$ and the Complexity Comparison in different CRs and Scenarios}
    \label{tab1}
    \resizebox{85mm}{34mm}{
    \begin{tabular}{c|c|c|c|c|c|c}
    \hline
    \multirow{3}{*}{CR}   & \multirow{3}{*}{Methods} & \multirow{2}{*}{Indoor} & \multirow{2}{*}{Outdoor} & \multicolumn{3}{c}{Complexity}                                                                           \\ \cline{5-7} 
                          &                          &                         &                          & \multirow{2}{*}{\begin{tabular}[c]{@{}c@{}}Trainable \\ Params \footnotemark[1] \end{tabular}} & \multicolumn{2}{c}{MACC} \\ \cline{3-4} \cline{6-7} 
                          &                          & NMSE                    & NMSE                     &                                                                              & Encoder      & Decoder     \\ \hline
    \multirow{4}{*}{1/4}  & CsiNet                   & -17.36                  & -8.75                    & 2.10M                                                                      & 1.09M                   & \textbf{4.39M}       \\
                          & CsiNet+                  & -27.37                  & -12.4                    & 2.12M                                                                          & 1.45M                    & 23.26M      \\
                          & FISTA                    & -10.46                  & -6.35                    & -                                                                            & \textbf{1.05M}        & 41.94M      \\
                          & FISTA-Net                & \textbf{-36.76}         & \textbf{-22.4}          & \textbf{1.09M}                                                               & \textbf{1.05M}        & 74.71M      \\ \hline
    \multirow{4}{*}{1/8}  & CsiNet                   & -12.7                  & -7.61                    & 1.05M                                                                        & 0.56M        & \textbf{3.86M}       \\
                          & CsiNet+                  & -18.29                  & -8.72                    & 1.07M                                                                          & 0.93M        & 22.73M      \\
                          & FISTA                    & -6.39                   & -2.91                    & -                                                                            & \textbf{0.52M}        & 20.97M      \\
                          & FISTA-Net                & \textbf{-26.5}                        & \textbf{-13.65}          & \textbf{0.56M}                                                                             & \textbf{0.52M}        & 53.74M      \\ \hline
    \multirow{4}{*}{1/16} & CsiNet                   & -8.65                   & -4.51                    & 0.53M                                                                        & 0.30M        & \textbf{3.60M}       \\
                          & CsiNet+                  & -14.14                  & -5.73                    & 0.55M                                                                        & 0.67M        & 22.47M      \\
                          & FISTA                    & -3.18                   & -1.15                    & -                                                                            & \textbf{0.26M}        & 10.49M      \\
                          & FISTA-Net                & \textbf{-17.51}          & \textbf{-7.57}           & \textbf{0.30M}                                                               & \textbf{0.26M}        & 43.26M      \\ \hline
    \multirow{4}{*}{1/32} & CsiNet                   & -6.24                   & -2.81                    & 0.27M                                                                        & 0.17M        & \textbf{3.47M}       \\
                          & CsiNet+                  & -10.43                  & -3.4                     & 0.29M                                                                        & 0.54M        & 22.34M      \\
                          & FISTA                    & -1.11                   & -0.35                    & -                                                                            & \textbf{0.13M}        & 5.24M       \\
                          & FISTA-Net                & \textbf{-12.01}         & \textbf{-4.41}           & \textbf{0.17M}                                                                        & \textbf{0.13M}        & 38.01M      \\ \hline
    \multirow{4}{*}{1/64} & CsiNet                   & -5.84                   & -1.93                    & 0.14M                                                                        & 0.11M        & 3.40M       \\
                          & CsiNet+ \footnotemark[2]                  & -5.99                   & -2.22                    & 0.16M                                                                        & 0.47M        & 22.27M      \\
                          & FISTA                    & -0.29                   & -0.05                    & -                                                                            & \textbf{0.07M}             & \textbf{2.62M}       \\
                          & FISTA-Net                & \textbf{-8.54}          & \textbf{-2.6}           & \textbf{0.10M}                                                                        & \textbf{0.07M}             & 35.39M      \\ \hline
    \end{tabular}}
    \end{table}
	\footnotetext[1]{There are no trainable parameters for FISTA.}
	\footnotetext[2]{The results of CsiNet+ under CR=$1/64$ are running on our platform due to \cite{8972904} does not show the detail results.}

	From Table \ref{tab1}, the FISTA-Net has better NMSE performance and lower complexity at all CRs in indoor and outdoor scenarios, especially when CR is high. Compared with FISTA, FISTA-Net has an significant performance boost in same iteration stages due to the powerful learning ability. On the other hand, FISTA-Net outperform CsiNet and its enhanced version, CsiNet+, demonstrating the advantage of model-driven deep learning methods. Meanwhile, all CSI feedback methods, including FISTA-Net, perform worse in the outdoor scenario than the indoor scenario due to the weaker sparsity and non-stationarity.
	
	As for the complexity, FISTA-Net also outperforms
	others on less trainable parameters and MACC in the encoder, which is vital for the deployment of the model on UE. Although the computational complexity of FISTA-Net in the decoder is no advantage, for which the higher MACC is mainly concentrated in gradient descent step in Eq. (11a), the decoder is deployed in BS with powerful computing power and inference ability. Besides, the technology of model compression and acceleration can be exploited to reduce the MACC before the deployment of the model.

	\section{Conclusion}
	In this paper, we designed a TSLR CSI feedback scheme for low rank downlink CSI feedback in FDD mmWave massive MIMO systems by a model-driven neural network, named FISTA-Net, which has an easy designing and interpretable structure and powerful learning ability. Furthermore, the ST-Net based attention mechanism to learn thresholds adaptively and learnable parameters have been applied in FISTA-Net to choose the optimal parameters to improve the performance. The simulation results showed the TSLR with FISTA-Net can reconstruct low rank CSI with high accuracy and lower complexity, and FISTA-Net also have excellent performance for other scenarios.

	% or
	

	\ifCLASSOPTIONcaptionsoff
	\newpage
	\fi

	\bibliographystyle{ieeetr}    %引用风格
	\bibliography{ref}		  %bib文件名，同时该语句确定了参考文献出现的位置
	
\end{document}